\documentclass[a4paper,11pt]{article}
\pdfoutput=1
\usepackage{jheppub}
\usepackage{amsmath,amssymb}
\usepackage{bbold}
\usepackage{graphicx}
\usepackage{float}
\usepackage{color}
\usepackage [utf8]{inputenc}
\usepackage{subfigure}
\usepackage{euscript}
\usepackage{placeins}
\usepackage{diagbox}
\usepackage[thinlines]{easytable}
\usepackage{ulem}
\usepackage{xcolor}

\allowdisplaybreaks[2] 

\def\OO{\mathcal{O}}
\def\bb{{\boldsymbol b}}
\def\bop{{\boldsymbol p}}

\def\bk{{\boldsymbol k}}

\def\bx{{\boldsymbol x}}

\def\gsqa{g_{3\mathrm{d}}^2a}

\def\gsql{g_{3\mathrm{d}}^2 L}
\def\gsqlmax{g_{3\mathrm{d}}^2 L_\mathrm{max}}
\def\gsqlmin{g_{3\mathrm{d}}^2 L_\mathrm{min}}
\def\gsq{g_{3\mathrm{d}}^2}
\def\gfour{g_{3\mathrm{d}}^4}
\def\gsix{g_{3\mathrm{d}}^6}
\def\gseven{g_{3\mathrm{d}}^7}
\def\geight{g_{3\mathrm{d}}^8}

\def\mD{m_{\mathrm{D}}}
\def\mDsq{m_{\mathrm{D}}^2}

\def\mIsq{m_{\infty}^2}

\def\Tr{\mathrm{Tr}}
\def\bp{{\bb_\perp}}

\def\Cbp{C(\bp)}

\def\EE{\Tr \, \langle E(L) U E(0) U^{-1} \rangle}
\def\BB{\Tr \, \langle B(L) U B(0) U^{-1} \rangle}
\def\xc{x_\mathrm{cont}}
\def\yc{y_\mathrm{cont}}

\def\Zg{Z_\mathrm{g}}
\def\Zgtd{Z_{\mathrm{g,3d}}}
\def\Zf{Z_\mathrm{f}}
\def\ZP{Z_\mathrm{P}}
\def\ZB{Z_\mathrm{B}}
\def\ZE{Z_\mathrm{E}}
\def\CR{C_\mathrm{R}}
\def\CA{C_\mathrm{A}}
\def\CF{C_\mathrm{F}}
\def\dA{d_\mathrm{A}}
\def\d{\mathrm{d}}
\def\e{\mathrm{e}}
\def\Eq#1{Eq.~(\ref{#1})}

\title{The nonperturbative contribution to asymptotic masses}
\author[a]{Guy D.\ Moore,} 
\author[a,b]{Niels Schlusser}
\affiliation[a]{Institut f\"ur Kernphysik, Technische Universit\"at Darmstadt\\
Schlossgartenstra{\ss}e 2, D-64289 Darmstadt, Germany}
\affiliation[b]{Department of Physics \& Helsinki Institute of Physics\\
P.O. Box 64, FI-00014 University of Helsinki}
\emailAdd{guy.moore@physik.tu-darmstadt.de, \\ nschlusser@theorie.ikp.physik.tu-darmstadt.de}

\abstract{
In gauge theories, charged particles obey modified
dispersion due to medium interactions (forward scattering), leading at
high energies $E \geq T$ to an asymptotic mass-squared $\mIsq$.
We calculate the infrared part of this mass nonperturbatively for the
theory of the strong interactions, QCD, through a lattice treatment of
its low-energy effective description, Electrostatic QCD (EQCD).
Incorporation of these results into a nonperturbative determination of
the effective thermal mass will require a still-incomplete
next-to-leading order perturbative matching of this quantity to full
QCD.
}

\keywords{quark-gluon plasma, dimensional reduction, effective
  theories, kinetic theory, lattice gauge theory}

\begin{document}

\maketitle

\section{Introduction}
\label{sec:intro}
The Quark-Gluon-Plasma (QGP) is an exotic state of matter created at
collider facilities
\cite{Arsene:2004fa,Back:2004je,Adams:2005dq,Adcox:2004mh} and
delivering important insights into the nature of strongly interacting
matter. The experimental effort to measure the features of the QGP
with the highest possible precision goes hand in hand with the
theoretical goal of giving precise predictions for these very
measurements.  The properties of the QGP can in essence be divided into
two classes:  thermodynamics and transport properties.
Typical examples for the former are the pressure, perturbatively
determined up to $\OO(g^6 \ln g)$
\cite{Shuryak:1977ut,Kapusta:1979fh,Toimela:1983407,Arnold:1994eb,Zhai:1995ac,Kajantie:2002wa}
and investigated nonperturbatively on the lattice
\cite{Bazavov:2014pvz,Borsanyi:2013bia}, and charge susceptibilities
and other quark-number correlations, which are well investigated on
the lattice \cite{Bazavov:2020bjn}.
Since thermal equilibrium is
time-translation invariant, these results can be derived in Euclidean
time, which simplifies the calculations considerably.
In contrast, the calculation of transport properties usually
requires an explicit treatment of Minkowski time, which is conceptually
more demanding. Nevertheless, there has been significant progress
in recent years, for instance for the shear viscosity
$\frac{\eta}{s}$ \cite{Meyer:2007ic,Ghiglieri:2018ltw} or the thermal
photon production rate \cite{Ce:2020tmx,Ghiglieri:2013gia}.

Anther interesting transport phenomenon to investigate is how the
medium modifies a \textsl{jet}, whose theoretical treatment is for
example summarized in \cite{Ghiglieri:2015ala}. For our purpose, jets
are strongly interacting particles that carry a large momentum and
move at (almost) the speed of
light.  Because of the large momentum scale, asymptotic freedom of QCD
suggests that the strong coupling constant $g$ is small
\cite{Gross:1973,Politzer:1973}, and that they can be well described
by an expansion in powers of $g$. In traversing through the medium,
however, they receive highly nontrivial modifications
\cite{Zakharov:1996fv,Zakharov:1998sv}, pictured by a large number of
soft bumps between jet and medium constituents. This interaction
eventually leads to \textsl{jet broadening}, i.e.\ a broadening of the
jet's transversal momentum distribution.
Related to that, \textsl{jet quenching} refers to a jet losing energy
to the medium \cite{Zakharov:1997uu,Baier:2000mf}, predominantly via
stimulated emission due to multiple soft scatterings with medium
constituents.

A quantity that influences both phenomena is the \textsl{asymptotic
  mass}. It was already found by Klimov and by Weldon that massless
particles, be it
gluons \cite{Klimov:1981ka,Weldon:1982aq}
or fermions \cite{Klimov:1982bv,Weldon:1982bn}, tend to
follow the dispersion relation of massive particles
\begin{equation}	\label{disp_rel}
\omega_\bop^2 = \bop^2 + m_\bop^2
\end{equation}
with their mass $m^2_\bop$ depending on the momentum $\bop$ and
generated by the interaction of the particle with the medium
(essentially, forward scattering). At large momenta, the particle's
velocity in the plasma frame approaches the speed of light
$v \to (1,\hat{\bop})$, and this thermal mass approaches a constant
which we will call the asymptotic mass $\mIsq$.
At the leading perturbative order, $\mIsq$ can
be computed from the imaginary part of the transverse
Hard-Thermal-Loop-resummed self-energy $\Pi_\mathrm{T}$ near the light
cone \cite{Braaten:1989kk,Braaten:1990it}.  The gluon's
asymptotic mass, for instance, reads \cite{Blaizot:2005wr}
\begin{equation}	\label{gluon_msq}
\mIsq \equiv \Pi_\mathrm{T}(k_0 = k) = \frac{\mDsq}{2} \, 
\end{equation}
to first perturbative approximation.  Here
$\mDsq = (2N_c+N_\mathrm{f}) g^2 T^2/6$ is the Debye screening mass squared.

The expression \eqref{gluon_msq} can be generalized to a particle in an
arbitrary representation $\mathrm{R}$ and related to a
combination of two condensates \cite{Braaten:1991gm,CaronHuot:2008uw},
a gauge contribution $\Zg$ and a fermionic contribution $\Zf$:
\begin{equation}	\label{def_asy_CH}
\mIsq = g^2 \CR \left( \Zg + \Zf \right) \, , 
\end{equation}
with the (running) gauge coupling $g$, and the Casimir operator of the
representation $\CR$ of the particle that makes up the jet. The two
condensates are \cite{Braaten:1991gm,CaronHuot:2008uw}
\begin{align}	\label{def_condensates}
  \Zf &\equiv \frac{1}{2 d_\mathrm{R}} \left\langle \overline{\psi}
  \frac{v_\mu \gamma^\mu}{v \cdot D} \psi \right\rangle \\
\Zg &\equiv -\frac{1}{g^2 \dA} \left\langle v_\sigma F^{\sigma \mu}
\frac{1}{(v \cdot D)^2} v_\rho {F^\rho}_\mu \right\rangle \notag \, ,
\end{align}
where $\dA=N_c^2-1$ is the dimension of the adjoint representation, in which
the gauge bosons live, and $d_\mathrm{R}$ is the dimension of the
representation $\mathrm{R}$ of the fermion, for instance the
fundamental representation of $SU(3)$ for quarks in QCD, and the
angular brackets represent an average over all thermal fluctuations.
Note that
our definition of the QCD field strength $F^{\sigma \mu}$ is the
geometrical one, $F^{\mu\nu} = i [ D^\mu,D^\nu ]$ with
$D^\mu = \partial^\mu -i A^\mu$; $g^2$ is absorbed into the
coefficient on the field strength in the action,
$S = \int \frac{1}{2g^2} \mathrm{Tr}\: F^2 + \ldots$.
These condensates represent the contribution of forward scattering in
the medium, with $1/ (v \cdot D)$ or $(v\cdot D)^2$ accounting for the
lightlike (eikonalized) propagation through the medium and
$\bar\psi,\psi$ or $v_\sigma F^{\sigma \mu}$ representing an
interaction with a medium excitation.  The leading-order,
long-distance approximation to these condensates reproduce
the forward-scattering part of the hard thermal loop
effective action along the light cone.

The gauge part of \eqref{def_asy_CH} can be related to an integral
over the correlator of two covariant Lorentz force vectors in position space \cite{Ghiglieri:2018ltw}
\begin{equation}	\label{def_int_condensate}
  \Zg = -\frac{1}{g^2 \dA} \int_0^\infty \hspace{-2pt}
  \d x_+ \,  x_+ \langle v_{k \, \mu} F^{\mu\nu}_a(x_+,0,0_\perp) 
  U^{ab}_\mathrm{A}(x_+,0,0_\perp; 0,0,0_\perp)
  v_{k \, \rho} F^{\rho}_{b \, \nu}(0,0,0_\perp) \rangle \, ,
\end{equation}
where the $x_+$ integration and the modified adjoint Wilson line
$U^{ab}_\mathrm{A}$ arise from $1/(v\cdot D)^2$.
Locations in space-time are described by
three-vectors $(x_+,x_-,x_\perp)$ in light-cone coordinates.
Because the jet is highly relativistic, we have
$x_-=t-z=0$ in the plasma frame. The problem
therefore effectively reduces to three dimensions, two in the
transverse plane $x_\perp$, where rotational invariance is present,
and a coordinate that encompasses the elapsed time $t$ for the jet as
well as its covered distance $z$ in the medium
$x_+ = (t + z)/2 \equiv L$.
We use the convention that the z-component of $\bk$ is the jet's
direction of propagation, so $v = (1, 0, 0, 1)$.

\begin{figure}[htbp!] 
\centering
\includegraphics[width=0.5\textwidth,keepaspectratio]{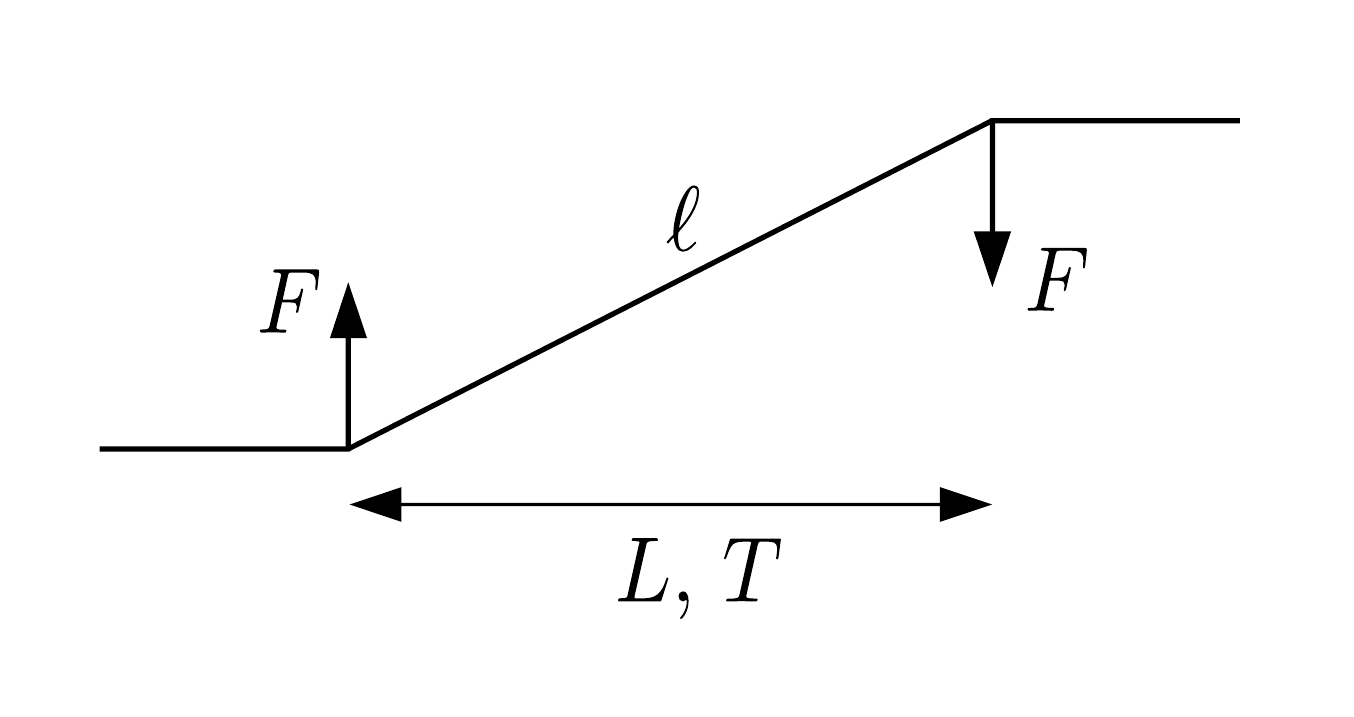}
\caption{Pictorial motivation for the form of the correlator
  \eqref{def_int_condensate}.}
\label{fig:cartoon}
\end{figure}

At this point, we would like to briefly build up some intuition for
the specific form of \eqref{def_int_condensate}. We can distinguish
two sorts of jet-medium interactions.  First, there is true
scattering, where the transverse momentum of the jet particle changes;
this is described by the transverse momentum
broadening kernel $\Cbp$ \cite{Zakharov:1998sv}.  But there is also
forward scattering, where the jet particle temporarily changes
direction before scattering back into its original state (or more
technically, a scattering creates an amplitude in a different momentum
state, which subsequently scatters again to return to the original
momentum, introducing a phase shift).  Such
forward scattering is what causes a dispersion correction.  It can be
understood qualitatively as the jet particle making a
scattering-induced detour, as pictured in
Fig.~\ref{fig:cartoon}.  The sum of all such forward scattering
possibilities induces an effective mass term \cite{Zakharov:1997uu}.

Relativistic kinematics tells us the length of the detour  
\begin{equation}
\ell^2 = L^2 + \frac{F^2 T^4}{m'^2 \gamma(v)} \equiv L^2 + \Delta \ell^2 \, ,
\end{equation}
where $m'$ is a dummy jet mass that will eventually drop out of the
calculation, $\gamma(v)$ is the gamma-factor of the jet at velocity
$v$, and $T$ is the time it takes for the jet to cover the distance
$L$. If we consider this problem from a frame of reference in which we
do not take into account the additional detour, the kinematic
invariant reads\footnote{We use the Minkowski-metric with the negative
  sign in the time-component.}
\begin{align}	\label{motivation}
p^\mu p_\mu = m'^2 \gamma^2(v) \left( -1 + \frac{\ell^2 - \Delta \ell^2}{T^2} \right) \xrightarrow{T \to \ell ,\, m' \to 0} &  - F^2 \ell^2 \approx - F^2 L^2
\end{align}
in the ultra-relativistic limit, in which also the contribution of longitudinal forces to the effective mass vanish. Consequently, the (transversal) gauge force must appear twice in the correlator resulting in the mass. Furthermore, a full quantum treatment requires a sum over all possible paths instead of just considering one as in Fig.~\ref{fig:cartoon}. Hence, our estimate  for the effective mass yields
\begin{equation}
\sum_L F^2 L^2 \rightarrow \int \hspace{-2pt} \d L \, L F^2 \sim \mIsq \, ,
\end{equation}
reflecting the qualitative features of \eqref{def_int_condensate}.

The state-of-the-art perturbative result for $\Zg$ at next-to-leading-order (NLO) \cite{CaronHuot:2008uw} is
\begin{equation}	\label{PT_Zg}
\Zg = Z_\mathrm{g}^\mathrm{LO} + \delta \Zg = \frac{T^2}{6} - \frac{T \mD}{2 \pi} \, .
\end{equation}
The Debye mass \cite{Weldon:1982aq}, in turn, depends on the coupling
$g$: in QCD ($N_c=3$) it reads
\begin{equation}
\mDsq = g^2 T^2 \left( 1 + \frac{N_\mathrm{f}}{6} \right) + \OO(g^4) \, .
\end{equation}
Formally the NLO correction is therefore suppressed by a single power
of the coupling $g$.  Numerically, it is not suppressed even at
$T=100$ GeV temperatures, indicating a very poor convergence for the
perturbative series.  This is a familiar situation for perturbative
calculations in hot QCD, and as usual, it arises because there are
large corrections arising from the large-occupancy, infrared sector of
QCD.  Specifically, the
convergence of the perturbative series is spoiled by the high
occupancy of the gluon 0-Matsubara-frequency mode.  When this mode
circulates in a loop, it introduces
an additional factor of $n_\mathrm{B} \sim 1/g$
\cite{Linde:1980ts,Appelquist:1981vg,Nadkarni:1982kb}. A possible
solution to this problem is provided by treating the badly-convergent
contributions separately in a framework called \textsl{Electrostatic Quantum Chromodynamics} (EQCD) \cite{Braaten:1995cm}.

In contrast, the
situation for $\Zf$ appears to be better under control. Fermions do
not feature a 0-mode so the above reasoning does not apply here.
Indeed, it was found that \cite{CaronHuot:2008uw}
\begin{equation}	\label{PT_Zf}
\Zf = Z_\mathrm{f}^\mathrm{LO} + \delta \Zf = \frac{T^2}{12} + 0 \, ,
\end{equation}
so the assumption of a convergent series for the fermionic part of
$\mIsq$ may be justified.

According to \cite{CaronHuot:2008ni}, correlators like
\eqref{def_condensates} can be computed in a setup that quantizes the
plasma on the light-cone. Their dynamics is dominated by multiple soft
scatterings that have to be resummed.  But because they are infrared,
they can be computed in
three dimensional EQCD instead of four dimensional full QCD in
real-time, with modified Wilson lines that account for the parallel
transport of the gauge configurations. In the past, this technique has
been predominantly used for the computation of the transverse
collision kernel $\Cbp$
\cite{CaronHuot:2008ni,Panero:2013pla,Moore:2019lgw}. Similarly to the
calculation of $\Cbp$, the EQCD operator agrees with its
full-QCD-equivalent only in the infrared (IR) regime where EQCD is
supposed to be a valid description of full QCD. In the ultraviolet
regime, however, a difference between the two operators will
emerge.  This difference should be under perturbative control and can
be included as part of a matching calculation between thermal QCD and
EQCD, when the matching is extended to include the operators of
\Eq{def_int_condensate}.

In the course of the present work, we will evaluate the soft operator
in EQCD and leave the purely perturbative matching calculation for
later investigation. As for $\Cbp$, perturbation theory, even in EQCD,
is not always sufficient to deal with the nonperturbative nature of
some high-temperature phenomena. Sometimes, it is necessary to solve
EQCD directly on the lattice, which has been done recently for $\Cbp$
\cite{Panero:2013pla,Moore:2019lgw}. This will also be our method of
choice for the determination of the soft part of asymptotic jet masses
in the following. 

We address the problem as follows. In Sec.~\ref{sec:formulation}, we
outline how we would compute $\Zg$ if there were no interactions. To
this end, we briefly introduce how our setup translates into EQCD,
show how the observables appear in this theory and finally present the free
solutions. Solving the full interacting theory requires discretizing
the theory and the corresponding observables on the lattice, done in
Section \ref{sec:lattice}. This part of the present work also outlines
how to deal with the contributions of \Eq{def_int_condensate} at the
shortest and longest distances, where the lattice determination
becomes problematic. Section
\ref{sec:results} involves a detailed presentation of our results
together with a thorough discussion and, where possible,
quantification of the errors that our results are fraught with. The
consequences of our results and how to improve on them is discussed in
Sec.~\ref{sec:conclusion}. Appendix \ref{app1} provides a detailed
overview of our lattice configurations.

\section{Formulation of the problem}
\label{sec:formulation}

\subsection{Electrostatic QCD}
Electrostatic Quantum Chromodynamics, or EQCD for short, is an
effective field theory for hot QCD.  At high temperatures, the
scaling behavior changes, for lengths longer than $1/T$ or energies
lower than $T$, to the behavior of a 3D theory, which is more strongly
coupled in the IR.  Within the Matsubara formalism for describing
thermal QCD, this arises because the usual frequency integral
$\int d\omega/2\pi$ is replaced by a discrete sum
$T \sum_{\omega=2n\pi T}$, with the $n=0$ (zero Matsubara frequency)
mode massless (at tree level), leading to an
$\OO(2\pi T / p)$ enhancement of small momentum $p$ parts of
diagrams.  For the electric sector this is cut off by the Debye mass,
but for the magnetic sector it is not cut off until magnetic degrees
of freedom become strongly coupled and confine.  For an overview
see Refs \cite{Linde:1980ts,Appelquist:1981vg,Nadkarni:1982kb}.
The idea of EQCD is to integrate out -- in a Wilsonian
renormalization-group-sense -- all Matsubara modes in QCD except for
these 0-modes, which can then be treated nonperturbatively.
Since there is only one Matsubara-mode left, one has
effectively eliminated the Euclidean time direction.  Therefore
the transition to EQCD is often called
\textsl{dimensional reduction}. A time direction that only consists of
one time layer does not allow for derivatives in time direction any
more. Consequently, this direction no longer features gauge
symmetry, which is what allows the temporal gauge field $A^0$, now
responsible for all electrical phenomena, to acquire a mass
$\mDsq$.  Loop level effects induce a self-coupling for this field,
denoted as $\lambda$. A scale is set by the now dimensionful
gauge coupling $\gsq = g^2 T$.  The full action of EQCD in the
continuum reads
\begin{equation} 
\label{cont_action}
S_{\mathrm{EQCD,c}} = \int \mathrm{d}^3x \, \left( \frac{1}{2 \gsq} \Tr \,
 F^{ij} F^{ij} + \Tr \, D^i \Phi D^i \Phi + \mD^2 \Tr \, \Phi^2 +
  \lambda \big( \Tr \, \Phi^2 \big)^2  \right) \, .
\end{equation}

The parameters in \eqref{cont_action} can be related to the full
theory via a perturbative matching procedure \cite{Braaten:1995cm}.
We employ a $\OO(g^4)$-accurate result \cite{Laine:2005ai}. Input parameters are the
temperature $T$ and the number of dynamical massless fermion flavors
$N_\mathrm{f}$ translating into the pair of (dimensionless) EQCD
parameters $x \equiv \lambda / \gsq$ and
$y \equiv \mDsq / \gfour \vert_{\mu=\gsq}$.  Note that $\mDsq$ varies
logarithmically with scale due to 2-loop effects; therefore in
defining $y$ we must specify the renormalization point.  Following
convention we choose $\mu=\gsq$.  We will consider the values given in
Tab.~\ref{match_scenarios}. EQCD also has a nontrivial phase structure
\cite{Kajantie:1998yc,Moore:2019lua}, with a transition between a
$\mathbb{Z}_3$-symmetric and broken phase that is of second order at
large $x$ and turns into a first order one for smaller $x$, including
all physically relevant $x$ values. Note the
$(x,y)$-pairs which describe full QCD prove to lie in the region where
the 3D theory prefers the $\mathbb{Z}_3$-broken phase.  Therefore we
are to consider metastable
$\mathbb{Z}_3$-symmetric states in a parameter range where the broken
phase is actually thermodynamically preferred.
Thus choosing sufficiently large volumes is crucial; they
keep the system from accidentally falling into the broken phase. It is
believed that dimensional reduction to EQCD is valid for all
$T \gtrsim 2 T_\mathrm{c}$ and possibly even further down
\cite{Laine:2003ay}, where $T_\mathrm{c}$ is the transition
temperature of full QCD.

\begin{table}[htbp!] 	
\centering
\begin{tabular}{|c|c|c|c|c|}	
\hline
$T$ & $n_{\mathrm{f}}$ & $x$ & $y$  \\
\hline
$250 \, \mathrm{MeV}$ & 3 & $0.08896$ & $0.452423$  \\
$500 \, \mathrm{MeV}$ & 3 & $0.0677528$ & $0.586204$ \\
$1 \, \mathrm{GeV}$ & 4 & $0.0463597$ & $0.823449$ \\
$100 \, \mathrm{GeV}$ & 5 & $0.0178626$ & $1.64668$ \\
\hline
\end{tabular}
\caption{EQCD parameters at four different temperatures and number of fermion flavors. }
\label{match_scenarios}
\end{table}

\subsection{Observables}
As already mentioned, one promotes \eqref{def_int_condensate} to EQCD
by replacing Wilson lines along the light cone with their modified
EQCD-counterparts \cite{CaronHuot:2008ni}
\begin{equation}
  \label{modifiedU}
\tilde{U}_\mathrm{R}(x_\perp,L;x_\perp,0) = \mathrm{P} \, \exp \left(
\int_0^L \hspace{-2pt} \d z (i A_z^a(x_\perp,z) + g_{3\mathrm{d}} \Phi^a(x_\perp,z))
T^a_\mathrm{R} \right) \, ,
\end{equation}
and $F^{i0}$ with $i D^i \Phi$ \footnote{In fact, it is $F^{i0} \to i \left[ D^i , \Phi \right]$, but we will omit the commutator for the sake of readability}. 
Applying rotational invariance in the transversal plane, we find
\begin{align}	\label{Zg_EQCD}
\Zgtd =& -\frac{2}{\dA} \int_0^\infty \hspace{-2pt} \d L \,  L \left(
- \langle \left( D^x \Phi(L) \right)_a \tilde{U}^{ab}_\mathrm{A}(L,0) \left( D^x
\Phi(0) \right)_b \rangle 
+ \langle F^{xz}_a(L) \tilde{U}^{ab}_\mathrm{A}(L,0)
F^{xz}_b(0) \rangle 
\right. \notag \\ & \phantom{-\frac{2}{\dA} \int_0^\infty \hspace{-2pt} \d L \,  L \bigg(} \left. 
+ \, i \, \langle (D^x \Phi(L))_a \tilde{U}_\mathrm{A}^{ab}(L,0) F^{xz}_b(0) \rangle 
+ i \, \langle F^{xz}_a(L)  \tilde{U}_\mathrm{A}^{ab}(L,0) (D^x \Phi(0))_b \rangle \right) \, . 
\footnotemark
\end{align}
\footnotetext{In an earlier version of this publication, we were not aware of the cross-correlations between $D_x \Phi$ and $F^{xz}$. They do not appear at tree-level; due to the $\Phi \to - \Phi$ symmetry in EQCD, an odd number of $\Phi$-field has to be sourced by the Wilson line. We thank Jacopo Ghiglieri for pointing that out to us.}
Using the modified Wilson loop \Eq{modifiedU} comes with the advantage of the correlator being free of logarithmically UV divergent perimeter law suppression which is cancelled by the scalar insertions $\Phi^a T^a$ in \Eq{modifiedU}. However, this correlator by itself is not free of $L \to 0$ UV divergences which have to be cancelled by a suitable subtraction. Since the correlators on the lattice can be measured at only finite $L$, anyway, we do \textsl{not} perform such a subtraction in this work and leave it to future investigations.

As a matter of choice, we work in the fundamental representation of
the correlators, related to the adjoint ones via
\begin{align}	
\langle \left( D^x \Phi(L) \right)_a \tilde{U}^{ab}_\mathrm{A}(L;0) \left( D^x
\Phi(0) \right)_b \rangle 
&= 2 \, \Tr \, \langle D^x \Phi(L)  \tilde{U}_\mathrm{F}(L;0) D^x \Phi(0)
\tilde{U}_\mathrm{F}^{-1}(L;0) \rangle \label{fund_EE_corr} \\ 
\langle F^{xz}_a(L) \tilde{U}^{ab}_\mathrm{A}(L;0) F^{xz}_b \rangle 
&= 2 \, \Tr \, \langle F^{xz}(L)  \tilde{U}_\mathrm{F}(L;0) F^{xz}(0)
\tilde{U}_\mathrm{F}^{-1}(L;0) \rangle \label{fund_BB_corr} \, ,
\end{align}
and for the cross-correlators, respectively.
In the following, we will drop the tilde on top of the modified Wilson 
lines $\tilde{U}$ for the sake of readability, and write half the right-hand 
side of \eqref{fund_EE_corr} as $\EE$ and half the right-hand side
of \eqref{fund_BB_corr} as $\BB$.

Let us stress again that the argument of \Eq{Zg_EQCD} agrees with that
of \Eq{def_int_condensate} only up to IR-safe terms which can and must
be determined in a matching calculation between full thermal QCD and
EQCD.  This calculation has not been carried out for these specific
operators, though it should proceed along similar lines to the
matching already conducted for $\Cbp$ in
\cite{CaronHuot:2008ni,Ghiglieri:2018ltw}.

Through insertions of the scale $\gsq$, everything can be recombined
into dimensionless quantities and we end up with the master formula
for our observable
\begin{align}	\label{Master_formula}
\frac{\Zg}{\gfour} =& 
\; \frac{1}{2}
\int_{\gsqlmin^\mathrm{EE}}^\infty \hspace{-2pt} \d \left( \gsql
\right) \, \gsql \, \frac{\Tr \, \langle D^x \Phi(\gsql)
  U_\mathrm{F}(\gsql;0) D^x \Phi(\gsql) U_\mathrm{F}^{-1}(\gsql;0)
  \rangle}{\gsix} 
  \notag \\ 
&- \frac{1}{2} \int_{\gsqlmin^\mathrm{BB}}^\infty \hspace{-2pt} \d
\left( \gsql \right) \, \gsql \, \frac{\Tr \, \langle F^{xz}(\gsql)
  U_\mathrm{F}(\gsql;0) F^{xz}(0) U_\mathrm{F}^{-1}(\gsql;0)
  \rangle}{\geight} 
\notag \\
&- \frac{i}{2} \int_{\gsqlmin^\mathrm{EB}}^\infty \hspace{-2pt} \d
\left( \gsql \right) \, \gsql \, \frac{\Tr \, \langle D^x \Phi(\gsql)
  U_\mathrm{F}(\gsql;0) F^{xz}(0) U_\mathrm{F}^{-1}(\gsql;0)
  \rangle}{\gseven} 
\notag \\
&- \frac{i}{2} \int_{\gsqlmin^\mathrm{EB}}^\infty \hspace{-2pt} \d
\left( \gsql \right) \, \gsql \, \frac{\Tr \, \langle F^{xz}(\gsql)
  U_\mathrm{F}(\gsql;0) D^x \Phi(0) U_\mathrm{F}^{-1}(\gsql;0)
  \rangle}{\gseven} \, ,
\end{align}
where $\gsqlmin^\mathrm{EE}$, $\gsqlmin^\mathrm{BB}$, and $\gsqlmin^\mathrm{EB}$ are minimal
separations at which we can measure the respective observable in the EQCD
effective picture on the lattice. Beyond that, we have to rely on
analytical solutions, as we will elaborate in Sec.~\ref{sec:lattice}. We will only calculate the $\EE$ and $\BB$ correlators in this work. These numerically dominate over the cross-correlations, anyway.

\subsection{Free solutions}
The short distance behavior of the full, continuum-extrapolated
correlators in \eqref{fund_EE_corr} and \eqref{fund_BB_corr} is
expected to match the perturbative expectations and eventually, at
sufficiently small separations, the free solutions. Since they not
only provide an important crosscheck for our data, but also will be of
practical use later, we will briefly present their instructive but
simple derivation in the following.

We assume the transverse direction to be oriented along the $x$-axis 
for simplicity. At tree-level, i.e.\ at $\OO(g_{3\mathrm{d}}^0)$, 
\eqref{fund_EE_corr} reduces to 
\begin{align}	\label{free_EE}
&\Tr \, \langle  D^x \Phi(L)  U_\mathrm{F}(L,0) D^x \Phi(L) U_\mathrm{F}^{-1}(L,0) \rangle \notag \\
& \hspace{80pt} = \left. \partial_x \partial_{x'} \, \Tr \, \langle \Phi(x,L) \Phi(x',0) \rangle \right|_{x,x' \to 0} \notag \\
& \hspace{80pt} = \CA \CF \, \partial_x \partial_{x'} \left( \frac{\e^{- \mD \, \sqrt{(x-x')^2 + L^2}}}{4 \pi \sqrt{(x - x')^2 + L^2}} \middle) \right|_{x,x' \to 0} \notag \\
& \hspace{80pt} = \frac{\e^{- \mD \, L}}{\pi L^3} \left( 1 + \mD \, L \right) \, ,
\end{align}
specializing to $SU(3)$ in the last step, and in a similar fashion for
\eqref{fund_BB_corr},
\begin{align}	\label{free_BB}
&\Tr \, \langle F^{xz}(L)  U_\mathrm{F}(L,0) F_{xz} U_\mathrm{F}^{-1}(L,0) \rangle \notag \\
& \hspace{80pt} = \left. \partial_x \partial_{x'} \, \Tr \, \langle A_z(x,L) A_z(0,0) \rangle \right|_{x \to 0} \notag \\
& \hspace{90pt} + \left. \partial_z \partial_{z'} \, \Tr \, \langle A_x(x,z) A_x(x',z') \rangle \right|_{x \to 0}^{z=L} \notag \\
& \hspace{80pt} = \frac{2 \CA \CF}{4 \pi L^3} - \frac{3 \CA \CF}{4 \pi L^3}  = - \frac{1}{\pi L^3} \, .
\end{align}
Plugging these results into \eqref{def_int_condensate}, 
we find a leading UV divergence of $\OO(L^{-3})$ which has to be subtracted. A contribution of $\OO(L^{-2})$ cancels at tree-level, whereas a $\OO(1/L)$-contribution persists, which when integrated over
$L \, \d L$ leads to a finite short-distance behavior in $\Zg$.
The superrenormalizable nature of EQCD means that, at the next loop
order, the small-$L$ behavior can be at worst $\OO(L^{-2})$.  However,
if there is no cancellation between the electric and magnetic
contributions at this order, there could be complications in the
short-distance behavior, leading potentially to logarithmic
short-distance contributions in \Eq{Zg_EQCD}.  If this is the case,
their role must be clarified by the matching to the 4D theory.

\section{Computational details}
\label{sec:lattice}

\subsection{Lattice implementation}
Our lattice implementation agrees with the one in
\cite{Moore:2019lgw}, based on a modification of Martin L\"uscher's
openQCD-1.6 \cite{openQCD}. Due to \cite{Moore:1997np} and
\cite{Moore:2019lua}, we are able to discretize the EQCD action on a
three-dimensional spatial lattice without any errors at $\OO(a)$. As
an update algorithm, we used a composite of 2 heatbath sweeps followed
by 4 overrelaxation sweeps through the volume in a
checkerboard-order. To improve the signal-to-noise ratio,
we applied the
multi-level algorithm \cite{Luscher:2001up,Meyer:2002cd}, dividing the
volume into four subvolumes along the $z$-axis and updating each
subvolume 80 times before a sweep through the
entire volume took place.

\subsection{Observables on the lattice and their continuum extrapolation}

For the $F^{xz}$-insertions in \eqref{fund_BB_corr}, we use the 
standard clover discretization of the field-strength tensor 
\cite{Weisz:1982zw,Weisz:1983bn}, with four 
$1\times1$ clover leaves.
The scalar field derivatives $D^x \Phi$ in
\eqref{fund_EE_corr} are discretized as
\begin{equation}
D^x \Phi(\bx) = \frac{1}{2a} \left( U(\bx) \Phi(\bx + a \boldsymbol
e_x) U^\dagger(\bx) - U^\dagger(\bx - a \boldsymbol e_x) \Phi(\bx - a
\boldsymbol e_x) U(\bx - a \boldsymbol e_x) \right) \, ,
\end{equation}
keeping in mind that the rescaling of the $\Phi$-fields to their
lattice equivalents yields another factor of $1/a$.  Note that the
clover magnetic field operator is spatially larger than the electric
field operator we use; therefore its $\OO(a^2)$ corrections are
larger, and we need more lattice spacings of separation between the
$B$ operators before continuum behavior sets in than for the $E$
operators.  Consequently we cannot pursue as short distances in the $\BB$
correlators as in the $\EE$ correlators.

All of our observables
are tree-level $\OO(a^2)$ accurate. Nevertheless,
there are still $\OO(a)$ discretization effects present generated by
loop effects in lattice perturbation theory. These generate
multiplicative $\OO(a)$-corrections, which we will denote by $\ZE$ and
$\ZB$ in the following.
For details about the discretization of the modified Wilson line, we
again refer to \cite{Moore:2019lgw} and \cite{Panero:2013pla}.
The  $\OO(a)$-renormalization of an isolated modified Wilson
line, or two well-separated modified Wilson lines, was calculated
analytically in \cite{DOnofrio:2014mld}.  We use the resulting
prescription, which is to scale $A^0$ in \Eq{modifiedU} by a $Z$
factor which is determined there.  However for the current case this
is insufficient, since there are new UV effects which can occur
because the two modified Wilson lines in \Eq{fund_EE_corr},
\Eq{fund_BB_corr} are coincident.  This will lead to new, so far
uncalculated, $\OO(a)$ perimeter-law renormalizations of the Wilson
line, which must be accounted for.  The overall lattice
expressions eventually read
\begin{align}
\frac{\EE_\mathrm{latt}}{\gsix a^3} =&  \left(
\frac{\EE_\mathrm{cont}}{\gsix} + \OO(a^2) \right) \,
\e^{\gfour\, a L \, \ZP}
\left( 1 + \gsqa\: \ZE \right),
\label{EE_cont_lim} \\ 
\frac{\BB_\mathrm{latt}}{\geight a^4} =&  \left(
\frac{\BB_\mathrm{cont}}{\geight} + \OO(a^2) \right) \,
\e^{\gfour\, a L \, \ZP}
\left( 1 + \gsqa \: \ZB \right).
\label{BB_cont_lim}
\end{align}

An analytical calculation of $\ZE$, $\ZB$, and $\ZP$ is probably
possible but appears to be technically somewhat complex, since the
Wilson line operator is an extended object.
However, by analyzing what diagrams could give rise to such linear in
$a$ effects, we find that the rescalings $\ZB$ and $\ZE$ do not depend
on the length $L$ (provided it is long in lattice units), and the
perimeter contribution $\ZP$ is strictly linear in $L$.  Furthermore,
the EQCD parameters $x$ and $y$ would first enter at a higher loop
order than the NLO effect giving rise to $\OO(a)$ corrections.
Therefore the $\ZE$, $\ZB$, and $\ZP$ factors are common to all
$(x,y)$ values and all separations.  They represent only 3
total fitting parameters over all of our data at multiple $(x,y)$ and
$L$ values.  Such global parameters make the fit more complicated,
introducing correlations between the fits to multiple lattice
spacings, separations, and $(x,y)$ pairs.  This made the fitting
procedure somewhat unstable, which we cured by using a
finite-difference solver to minimize $\chi^2$.  The correlations
between all datapoints must also be considered when performing the
final numerical integrations.  But because we only lose fitting 3
degrees of freedom over all data, the impact on the overall error
budget is relatively modest.

\begin{figure}[htbp!] 
\centering
\subfigure[$\frac{\EE}{\gsix}$]{\includegraphics[scale=0.55]{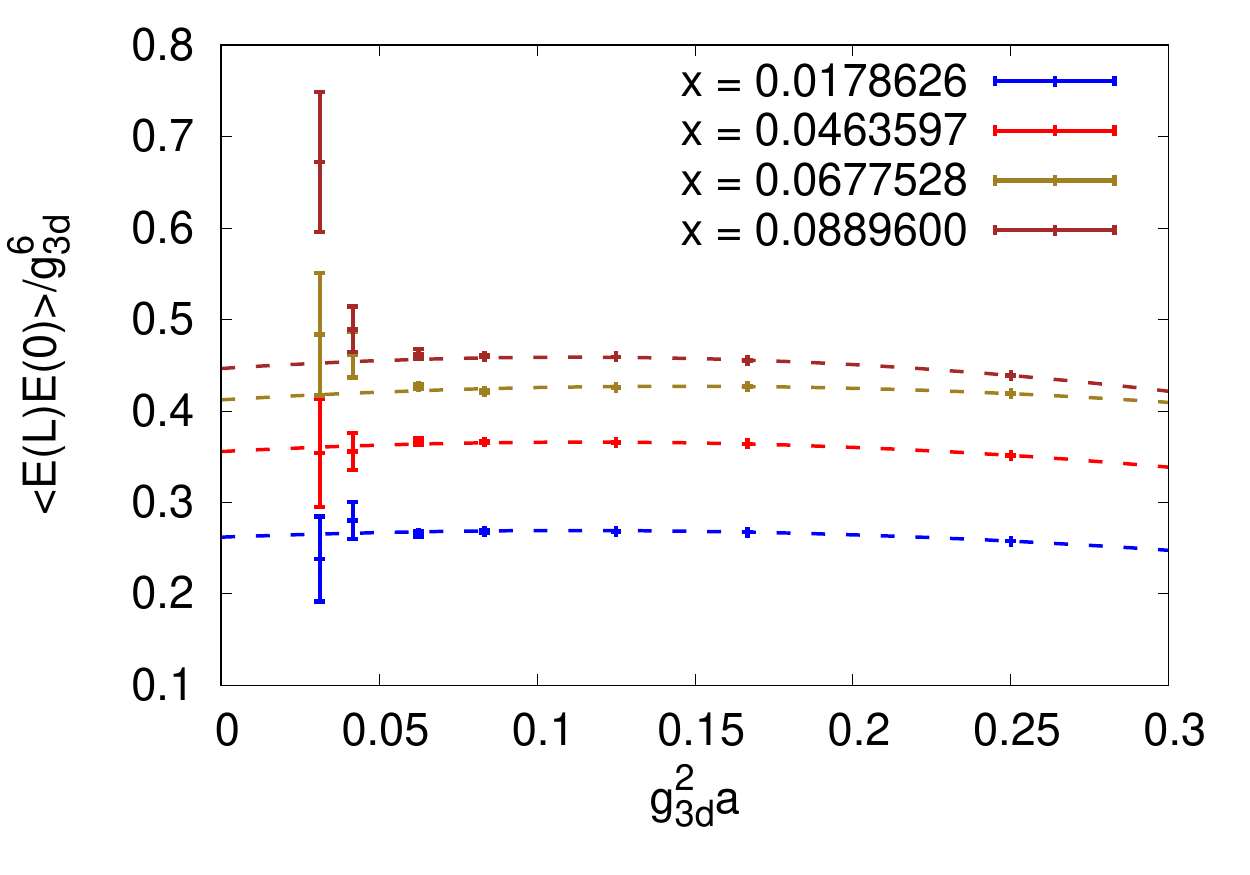}} 
\subfigure[$\frac{\BB}{\geight}$]{\includegraphics[scale=0.55]{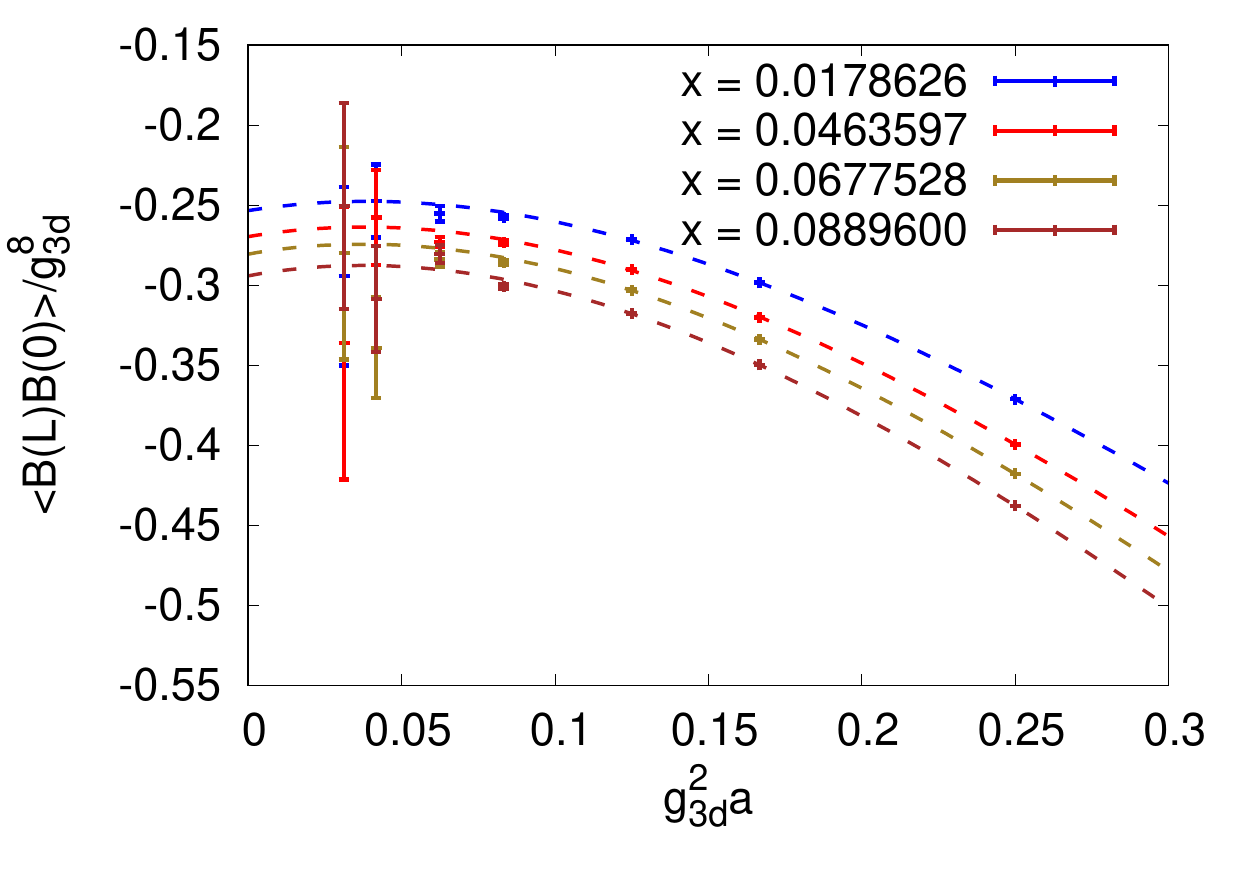}} 
\caption{Two continuum limits for $\gsql=1.0$ at all four considered
  temperatures}
\label{fig:plot_cont_lim}
\end{figure}

Fig.~\ref{fig:plot_cont_lim} shows a generic example for the continuum
extrapolation of the two correlators at a given length $\gsql = 1.0$.
We clearly see that $\OO(a)$-contributions
are still present, but the consideration of parabolae with joint
linear coefficients according to \eqref{EE_cont_lim} and
\eqref{BB_cont_lim} results in curves that fit our data reasonably
well. The obtained continuum values and their respective quadratic
coefficients are close to each other, but can still be
distinguished.

We considered four $(x,y)$ pairs, each at 8 separations $L$
for the $\EE$ correlator and 7 separations for the $\BB$ correlator,
and at 7 different lattice spacings.  (Note that not every
lattice spacing is useful for each length $L$; and we cannot evaluate
the $\BB$ correlator at the smallest $L$ value because as noted before,
the clover $B$ operator is rather large, requiring larger $L/a$ values
before one enters the continuum regime.)  For each $(x,y)$ pair
and separation, the $\EE$ and $\BB$ correlator are each continuum
extrapolated, fitting to a constant and quadratic term, plus linear terms
determined by the three global parameters $\ZE$, $\ZB$, and $\ZP$.
The likelihood of the grand continuum extrapolation
was reasonable with a $\chi^2 \approx 270$ on $276 - 123 = 153$
degrees of freedom.

\subsection{Integration of lattice data}
Even without the contributions from the 4D to 3D matching calculation,
we can still attempt to perform the integral shown in
\Eq{Master_formula}, at least for $L$ above some small-$L$ cutoff.
Our lattice data is available only within a finite range of
separations $\gsql$. We integrate this data using the trapezoidal
rule.  Ideally we would want data at a densely spaced set of distances
$L$, but this is impractical because the $L$ spacing is limited by the
lattice spacings of the coarsest lattices used.  Also, using
denser-spaced $L$ values would not actually help very much, because
the different $L$-value data are from the same simulations, and data
at nearby separations $L$ are highly correlated.
Nevertheless, the finite spacing in $\gsql$
introduces another error in $\Zg$ which we have to bear in mind. We
choose to measure $\EE$ at
$\gsql = 0.25, \, 0.5, \, 0.75, \, 1.0, \,1.5, \, 2.0, \, 2.5, \, 3.0$
and $\BB$ at
$\gsql = 0.5, \, 0.75, \, 1.0, \, 1.5, \, 2.0, \, 2.5, \, 3.0$,
where we elaborate on this choice in App.~\ref{app1}. We will
eventually find that using the trapezoid rule only 
introduces a subdominant discrete-integration-error.

\subsection{Integration of the tails}
The correlator \eqref{def_int_condensate} requires an integration of
the condensate up to infinite separations. Since the signal-to-noise
ratio drops dramatically with increasing $L$, correlators at
separations beyond $\gsql \geq 3.0$ are not determined robustly.
Nevertheless, knowing the analytical form of the
correlators at large separations $L$, we can fit the tail to the
largest data we have available. Similar things have been done, for
instance, in the
context of the hadronic contribution to the muon $g-2$-factor
\cite{Gerardin:2019rua}. However, their considerations cannot be
precisely applied in our case since we do not know the exact form of
the higher excited states.
For the magnetic contribution, Laine and Philipsen \cite{Laine:1997nq}
conjecture a functional form
\begin{equation}	\label{model_tail}
\Tr \, \left\langle F^{xz}(L)  U_\mathrm{F}(L,0) F_{xz}
U_\mathrm{F}^{-1}(L,0) \right\rangle = \frac{A}{\gfour L^2} \e^{-B \,
  \gsql} \, .
\end{equation}
Even though they consider a conventional Wilson line connecting the
two $B$-fields, this form is still applicable.  However in our case
the coefficient $B$ has a valid continuum limit, rather than having
logarithmically UV divergent perimeter contributions.  Each
coefficient is nonperturbative; we cannot predict them.

Based on similar physical reasoning, and
consistent with the free solution \eqref{free_EE}, we expect
the functional
form of the large-separation limit of $\EE$ also to be
\eqref{model_tail}, with different coefficients to be determined from
the fit. We choose our fitting range for the
large-$\gsql$-tail to be $ \left[ 1.5 ,\, 3.0 \right]$. This is hard
to justify a priori, since the correlation length above which the
lower bound of the interval must be is precisely $\xi \sim 1 /
m_\infty$. However, we will eventually find that $1.5 \gg \gsq /
m_\infty$.

At the other end, we do not have data at separations below certain
minimal separations $\gsqlmin^\mathrm{E}$ and
$\gsqlmin^\mathrm{B}$. Below these minimal separations, one would have
to connect to an EQCD perturbative expression. Since the (yet unknown)
difference between the EQCD effective description and the ``true''
full-QCD version of \eqref{def_int_condensate} is expected to
contribute significantly there, we leave the consideration of this
region entirely for future study.

\section{Results}
\label{sec:results}
Before we give our numerical results, we present a thorough discussion
of the multiple possible sources of both statistical and systematic
errors throughout our calculation. As numerical results, we provide
continuum-extrapolated data for $\EE$ and $\BB$ in
Figs.~\ref{fig:cont_EE} and \ref{fig:cont_BB}, the corresponding
values can be found in Tab.~\ref{sim_res_tab}, and details about the
simulations appear in Tab.~\ref{sim_params} of App.~\ref{app1}.

\subsection{Error estimation}
The estimation of the error has to take several sources into account. They are:
\begin{enumerate}
\item The statistical fluctuations of our data induced by the Monte-Carlo simulation of the path-integral
\item The uncertainty in choosing the model for the long-$L$-tail and in the determination of its parameters
\item The discretized evaluation of the $L\, \d L$-integration
\item The matching to the perturbative solutions at small $\gsql$
\item The reliability of reducing full QCD to EQCD
\item The perturbative determination of $\Zf$
\end{enumerate}

We will now discuss these aspects in detail, and give and explain
quantitative estimates where possible, see Tab.~\ref{sim_res_tab}.

The Monte-Carlo error of $\EE$ and $\BB$ can be determined by a
standard binning and Jackknife analysis. Since the
continuum extrapolation was performed in a grand fit, one has to
account for correlations of all lattices at all separations
$\gsql$. Even if their $\gsqa$ naively would not contribute, it does
so via its influence on the values of $\ZP$, $\ZB$, and $\ZE$. This
error, $\Delta_\mathrm{MC}$, is given in the first brackets in
Tab.~\ref{sim_res_tab}.

Another dominantly statistical error is created when we match the
tails of our models for $\EE$ and $\BB$, see \eqref{model_tail}. To be
precise, the origin of this error is twofold; the assumption of the
model itself introduces a systematic error, and the determination of
the model parameters from our data at large $\gsql$ generates a
statistical error. A valid estimate for this is provided by
determining the coefficients in \eqref{model_tail} by a fit,
performing the integral, repeating this procedure for each Jackknife
bin, and running the standard Jackknife-analysis over the different
values for the tail-integral.  This procedure includes the
correlations between the fitting parameters in \Eq{model_tail}.
In practice, we found that the
statistical error amounts to about 100\% of the correction induced by
the large-$\gsql$-tail, which seems to be a reasonable estimate of the
overall error of the large-$\gsql$-integration. This error,
$\Delta_\mathrm{LT}$, is given in the second brackets of
Table~\ref{sim_res_tab}.

Our data in the available window was integrated using the trapezoidal rule. The remainder $\Delta_\mathrm{TR}$ of such a discretized integration bounded by $\gsqlmin$ and $\gsqlmax$ reads
\begin{equation}	\label{trapezoidal_error}
\Delta_\mathrm{TR} = \frac{\gsqlmax - \gsqlmin}{12} \, \max_{\gsql} \, | I''(\gsql)| \, ,  
\end{equation}  
where the maximum of the second derivative of the integrand $I$ in
\eqref{Zg_EQCD} within the interval $\left[ \gsqlmin,\, \gsqlmax
  \right]$ has to be found, involving the additional factor of $\gsql$
from the $L \, \d L$-integration. This is tricky since the overall
functional form of our data is precisely what we do not know. However,
we know the functional form of the tails which we take as a
basis. This might not be a rigorous choice but we will see that this
error is in any case subdominant compared to other error sources.
For this consideration, we extend the range of validity of the
free solutions up to $\gsql=1.0$ and the range of validity of the
large-$\gsql$-tails down to the same value. Consequently, we have
three trapezoids of length $\delta (\gsql)=0.25$ with free-solution
estimates of the second derivative and four trapezoids of length
$\delta (\gsql)=0.5$ with tail estimates. We give this systematic
error in the third brackets in Tab.~\ref{sim_res_tab}.



Furthermore, the dimensional reduction to EQCD also introduces errors.
These errors are of two types.  First, there is the precision with
which the matching is computed for the Lagrangian of EQCD.  These
corrections are formally suppressed by $\OO(g^4)$ relative to the
leading behavior.  Second, there is the
accuracy of the matching for the specific operator we are
considering.  So far this has only been performed at tree level, which
as we have already noted is insufficient.  A calculation at the next
order would represent parametrically $\OO(g^4)$ contributions to
$\Zg$, which is formally $\OO(g^2)$ relative to the leading behavior.
Without this matching there are
formally $\OO(g^2)$ relative errors from the matching calculation;
with an NLO calculation this would be reduced to formally $\OO(g^4)$
errors.  The reason an NLO matching is needed is that nonperturbative
EQCD effects are also formally of relative order $\OO(g^2)$; therefore
without an NLO matching calculation for the operator, we have not
improved the formal order of the uncertainty.  It is difficult to
estimate the size of these (systematic) matching errors.  An NLO
calculation would shed considerable light on this point.

Last but not least, the current perturbative determination of $\Zf$
neglects terms suppressed by a factor of $g^2$.  These could be
determined in an NLO calculation of this condensate within the full 4D
theory.

All in all, the values of $\Zg$ in Tab.~\ref{sim_res_tab} are of the form
\begin{equation}
\text{best estimate}(\Delta_\mathrm{MC})(\Delta_\mathrm{LT})(\Delta_\mathrm{TR}) \, .
\end{equation}
Here $\Delta_\mathrm{MC}$, $\Delta_{\mathrm{LT}}$, and
$\Delta_\mathrm{TR}$ are respectively the errors from Monte-Carlo statistics
including the continuum extrapolation, the long-distance tail, and the
use of the trapezoid rule.

The overall quantified error of our results is dominated by
integration of the long-distance tail. This introduces a relative
error of $\sim 1\%$. However, we assume that analytical uncertainties
like the perturbative matching of the action and the operator and the
perturbative determination of $\Zf$ would not allow for higher
precision anyway, so an improvement is not mandatory.

\subsection{Numerical results}

\begin{figure}[htbp!] 
\centering
\includegraphics[width=\textwidth,keepaspectratio]{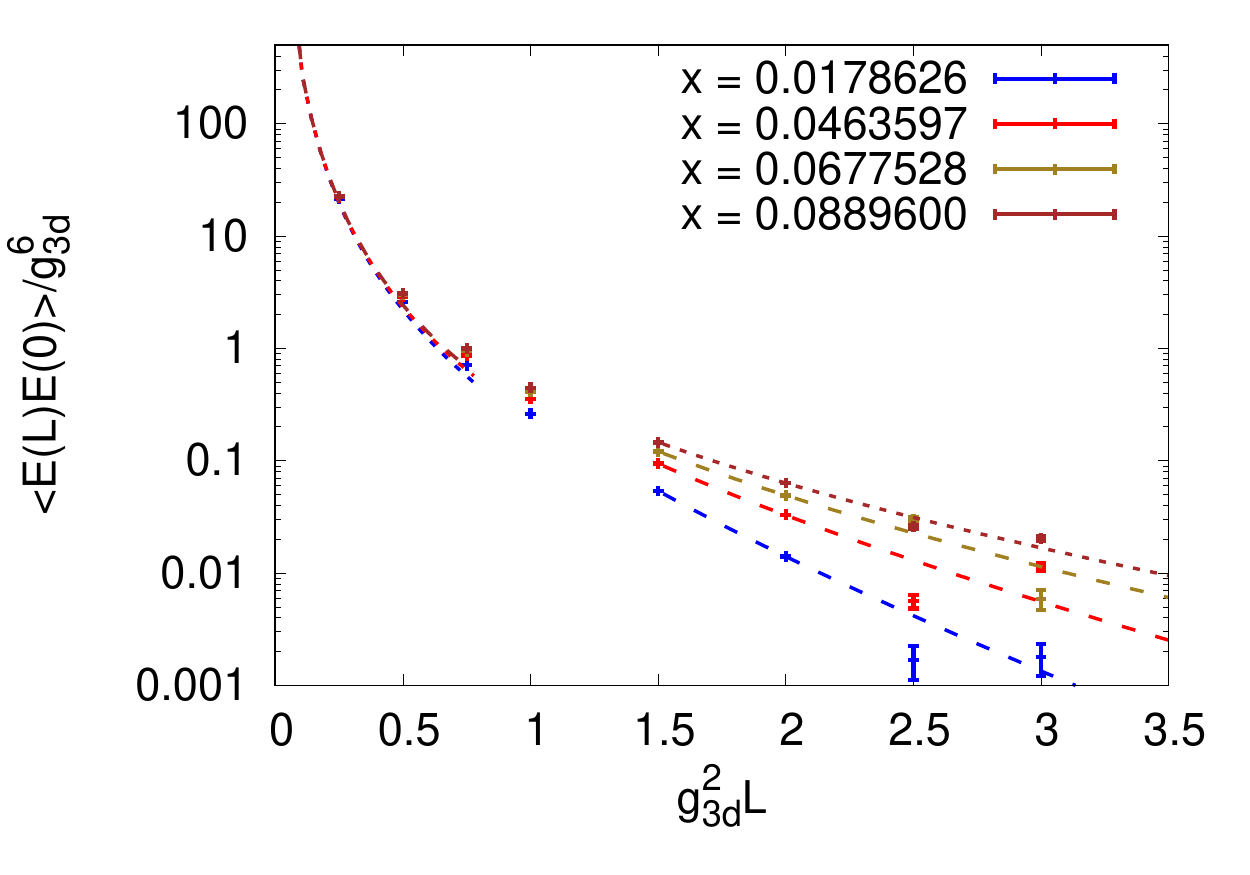}
\caption{Continuum-extrapolated version of $\frac{\EE}{\gsix}$,
  connecting to the free solution \eqref{free_EE} at small $\gsql$ and
  to an exponential tail of the form \eqref{model_tail} with the
  coefficients determined from our data at $\gsql \leq 1.5$. The
  short-dashed lines indicate the respective free solutions
  corresponding to a pair of $x$ and $y$. The long-dashed lines refer
  to the respective fitted tail function of the form
  \eqref{model_tail}.}
\label{fig:cont_EE}
\end{figure}

\begin{figure}[htbp!] 
\centering
\includegraphics[width=\textwidth,keepaspectratio]{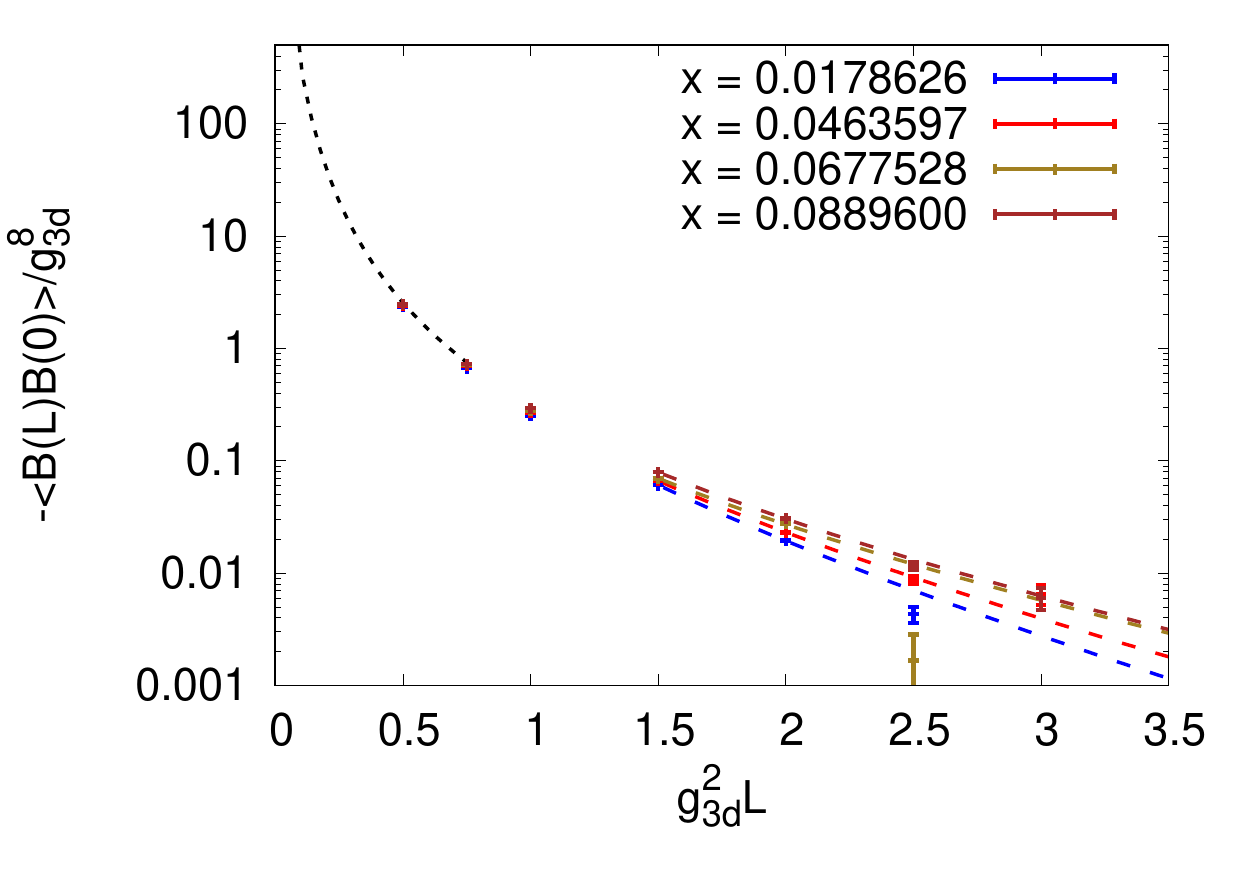}
\caption{Continuum-extrapolated version of $- \frac{\BB}{\geight}$,
  connecting to the free solution \eqref{free_BB} at small $\gsql$ and
  to an exponential tail of the form \eqref{model_tail} with the
  coefficients determined from our data at $\gsql \leq 1.5$. The black
  dashed line indicates the free solution that is independent of $x$
  and $y$. The dashed lines in color refer to the respective fitted
  tail function of the form \eqref{model_tail}.}
\label{fig:cont_BB}
\end{figure}

The plots Figs.~\ref{fig:cont_EE} and \ref{fig:cont_BB} show agreement
of our continuum data with the free solutions from \eqref{free_EE} and
\eqref{free_BB} at small $\gsql$. Take note that, in contrast to
$\EE$, we do not have data for $\BB$ at $\gsql=0.25$ since the
discretization effects are more severe for this operator and we did
not compute fine enough lattices to perform a valid continuum
extrapolation at this distance. 
At the other end of the range, we fit the functional form
\eqref{model_tail} to the correlators through the last four
points. The fit is dominated by the two innermost of these four points
since the signal-to-noise-ratio drops dramatically at increasing
$\gsql$. In turn, this phenomenon leads to some of the points in
Fig.~\ref{fig:cont_BB} at $\gsql = 3.0$ even being negative and therefore
not being displayed in the logarithmically scaled plot. As we can see
from their values in Tab.~\ref{sim_res_tab}, they are still consistent
with 0 within a few standard deviations.

We see in both plots that the dependence on the EQCD parameters ($x$
and $y$ from Tab.\ref{match_scenarios}) diminishes in going to small
$\gsql$. For $\BB$ in Fig.~\ref{fig:cont_BB}, this becomes evident by
looking at the free solution \eqref{free_BB} which has neither a dependence
on $x$ nor on $y$, and is jointly approximated by all four scenarios. For
Fig.~\ref{fig:cont_EE} displaying $\EE$, the situation is less
obvious. Expanding the exponential in \eqref{free_EE} for small
$\mD \, L$, we see that the dependence on $\mD$, or $\sqrt{y}$
respectively, comes with a suppression of $\OO(L^2)$ and therefore
becomes negligible at small distances.

In both cases, we observe a strict hierarchy: the correlators $\EE$ and $-\BB$
shrink with increasing $x$ (corresponding to decreasing the
temperature).  We also find that $\EE$ and $-\BB$ become closer to
each other as one increases the temperature, which is expected due to
the screening mass $\mD$ increasing with temperature and reducing
scalar correlations over large distances.

Adding $\EE$ and $-\BB$ and integrating according to
\eqref{Master_formula}, as discussed in the previous section, one indeed obtains a strictly positive integrand, yielding a positive $\Zg$.

Since data for the cross-correlations and the NLO matching are still missing, we postpone the calculation of $\Zg$ to a later publication.

\begin{table}[ptbh] 
  \centering {\small
    \begin{tabular}{|c|c|c|c|c|}
      \hline
      &
      \multicolumn{2}{c|}{$x=0.08896$}
      &
      \multicolumn{2}{c|}{$x=0.0677528$}
      \\
      &
      \multicolumn{2}{c|}{$y=0.452423$}
      &
      \multicolumn{2}{c|}{$y=0.586204$}
      \\
      &
      \multicolumn{2}{c|}{$N_f=3,\; 250\:\mbox{MeV}$}
      &
      \multicolumn{2}{c|}{$N_f=3,\; 500\:\mbox{MeV}$}
      \\
      \hline
      $\gsql$ & $\langle EE \rangle$ & $\langle BB \rangle$ 
      & $\langle EE \rangle$ & $\langle BB \rangle$ \\ \hline
      $0.25$ & $22.761(41)$ & - & $22.559(35)$ & - \\
$0.5$ & $3.1275(60)$ & $-2.5672(98)$ & $3.0230(61)$ & $-2.519(10)$ \\
$0.75$ & $1.0116(32)$ & $-0.756(11)$ & $0.9436(30)$ & $ -0.7349(91)$ \\
$1.0$ & $0.4539(25)$ & $-0.3217(33)$ & $0.4092(19)$ & $-0.3057(33)$ \\
$1.5$ & $0.1470(19)$ & $-0.0995(51)$ & $0.1186(20)$ & $ -0.0822(40)$ \\
$2.0$ & $0.0713(31)$ & $-0.0255(72)$ & $0.0448(19)$ & $-0.0217(56)$ \\
$2.5$ & $0.033(13)$ & $0.005(42)$ & $0.020(11)$ & $0.027(30)$ \\
$3.0$ & $0.036(20)$ & $0.034(46)$ & $0.023(17)$ & $0.060(56)$ \\
 	  \hline
      \hline
      &
      \multicolumn{2}{c|}{$x=0.0463597$}
      &
      \multicolumn{2}{c|}{$x=0.0178626$}
      \\
      &
      \multicolumn{2}{c|}{$y=0.823449$}
      &
      \multicolumn{2}{c|}{$y=1.64668$}
      \\
      &
      \multicolumn{2}{c|}{$N_f=4,\; 1\:\mbox{GeV}$}
      &
      \multicolumn{2}{c|}{$N_f=5,\; 100\:\mbox{GeV}$}
      \\
      \hline
      $\gsql$ & $\langle EE \rangle$ & $\langle BB \rangle$
      & $\langle EE \rangle$ & $\langle BB \rangle$ \\ 
      \hline
      $0.25$ & $22.290(33)$ & - & $21.588(34)$ & - \\
$0.5$ & $2.8991(48)$ & $-2.4833(88)$ & $2.6111(59)$ & $-2.4214(88)$ \\
$0.75$ & $0.8686(22)$ & $-0.7150(69)$ & $0.7107(35)$ & $-0.673(10)$ \\
$1.0$ & $0.3574(18)$ & $-0.2922(30)$ & $0.2616(20)$ & $-0.2750(30)$ \\
$1.5$ & $0.0957(15)$ & $-0.0831(50)$ & $0.0530(12)$ & $-0.0580(32)$ \\
$2.0$ & $0.0317(21)$ & $-0.02885(59)$ & $0.0133(16)$ & $-0.0232(38)$ \\
$2.5$ & $0.0179(93)$ & $-0.033(33)$ & $-0.0020(59)$ & $-0.006(25)$ \\
$3.0$ & $0.0223(13)$ & $-0.010(35)$ & $-0.0012(64)$ & $-0.024(20)$ \\
	  \hline
    \end{tabular}
}
\caption{%
  Results for the correlators $\EE / \gsix$ and $\BB / \geight$ at four temperatures over a range of separations. 
}
\label{sim_res_tab}
\end{table}

\section{Conclusion and outlook}
\label{sec:conclusion}

In the present work, we have shown how to compute the asymptotic
masses $\mIsq$ in QCD at high temperature nonperturbatively.
We showed that the real-time continuum expression,
\Eq{def_condensates} and \Eq{def_int_condensate},
can be matched to EQCD operators, \Eq{Zg_EQCD}, to compute the crucial
nonperturbative infrared contributions.  We then evaluated these
operators numerically within EQCD on a 3D lattice, performing a
precise extrapolation to the continuum limit.  Our data are presented
in Table \ref{sim_res_tab}, and we discussed a procedure for
numerically integrating these data to determine the IR contribution to
the asymptotic masses, and for assessing all sources of statistical
and systematic error in using the EQCD data.

Unfortunately, we find that to use the numerical results properly, we
need to complete the matching calculation for the operators of
\Eq{def_int_condensate} between full QCD and EQCD a next-to-leading
order in the perturbative expansion.  Therefore we are in the unusual
situation that the numerical study of EQCD is currently ahead of the
analytical understanding of the matching calculation.  Such a matching
calculation would improve the short-distance part of the EQCD
calculation by providing an NLO result for the small-separation
correlation functions, and it would ensure that all effects up to the
desired $\mIsq \sim g^4 T^2$ level of accuracy are accounted for.
In terms of future improvements, this is clearly the most pressing,
since without this matching calculation our results cannot yet be
used.  Further improvements to our numerical calculation are possible,
in particular by sharpening the error bars and computing down to
shorter distances, but generally the data appears to be in good shape
and the matching calculation is the most urgent need.

Together with the transverse collision kernel $\Cbp$ from
\cite{Moore:2019lua}, the then-complete-$\mIsq$ will hopefully
contribute significantly to understanding how a jet is modified by the
medium that it traverses. We are eager to see the impact of both
results on phenomenological transport calculations and how they match
experimental observations.

\section*{Acknowledgments}
This work was supported by the Deutsche Forschungsgemeinschaft (DFG,
German Research Foundation) – project number 315477589 – CRC TRR
211. Calculations for this research were conducted on the Lichtenberg
high performance computer of the TU Darmstadt. N. S. acknowledges 
support from Academy of Finland grants 267286 and 320123. We thank 
S\"oren Schlichting and Ismail Soudi for fruitful discussions on related 
topics in the course of which the idea for this project was born.
Last but not least, we would like to express our gratitude to Jacopo Ghiglieri, for pointing out the non-vanishing $\langle D^x \Phi U F^{xz} \rangle$ cross-correlations.

\appendix
\section{Simulation parameters}
\label{app1}

The parameters of our simulations can be found in
Tab.~\ref{sim_params}. The volumes were chosen large enough such that
the suppression of finite-volume effects according to
\cite{Hietanen:2008tv} was maintained. All separations of both $\EE$
and $\BB$ were measured on the same lattice configurations,
introducing correlations between the different $\gsql$ that had to be
accounted for in the error analysis.
The measurement of the very smallest separation, $\gsql = 0.125$ on
the $\gsqa = 1/32$-lattices could not contribute to any continuum
extrapolation since data was available at that particular spacing
only. We used this data just for crosschecking. The statistics for the
$T = 250 \, \mathrm{MeV}$-lattices was mostly higher since we used
this case to investigate how large separations $\gsql$ could be taken into
account. On each of the other temperatures, about $12000 \,
\mathrm{CPU} \cdot \mathrm{hrs}$ on Intel-Xeon E5-2670-processors were
spent.

\begin{table}[htbp!] 
\centering {\footnotesize
\begin{tabular}{|c|c|c|c|c|c|}	
\hline
$\gsqa$ & $\xc$ & $\yc$ & $N_\mathrm{x} N_\mathrm{y}
N_\mathrm{z}$
& $L / a$ & statistics\\
\hline
$1/4$ & $0.08896$ & $0.452423$ & $24^2 \times 48$ & $4,6,8,10,12$ & $94560$ \\
$1/6$ & $0.08896$ & $0.452423$ & $36^2 \times 72$ & $6,12,18$ & $88740$ \\
$1/8$ & $0.08896$ & $0.452423$ & $48^2 \times 96$ & $4,6,8,12,16,20,24$ & $61420$ \\
$1/12$ & $0.08896$ & $0.452423$ & $72^2 \times 144$ & $6,12,18,24,30,36$ & $8820$ \\
$1/16$ & $0.08896$ & $0.452423$ & $96^2 \times 192$ & $4,8,12,16,24,32,40$ & $2400$ \\
$1/24$ & $0.08896$ & $0.452423$ & $144^2 \times 288$ & $6,12,18,24$ & $480$ \\
$1/32$ & $0.08896$ & $0.452423$ & $192^2 \times 384$ & $4,8,16,24,32$ & $200$ \\
\hline
$1/4$ & $0.0677528$ & $0.586204$ & $24^2 \times 48$ & $4,6,8,10,12$ & $50200$ \\
$1/6$ & $0.0677528$ & $0.586204$ & $36^2 \times 72$ & $6,12,18$ & $29480$ \\
$1/8$ & $0.0677528$ & $0.586204$ & $48^2 \times 96$ & $4,6,8,12,16,20,24$ & $23940$ \\
$1/12$ & $0.0677528$ & $0.586204$ & $72^2 \times 144$ & $6,12,18,24,30,36$ & $8680$ \\
$1/16$ & $0.0677528$ & $0.586204$ & $96^2 \times 192$ & $4,8,12,16,24,32,40$ & $3600$ \\
$1/24$ & $0.0677528$ & $0.586204$ & $144^2 \times 288$ & $6,12,18,24$ & $480$ \\
$1/32$ & $0.0677528$ & $0.586204$ & $192^2 \times 384$ & $4,8,16,24,32$ & $360$ \\
\hline
$1/4$ & $0.0463597$ & $0.823449$ & $24^2 \times 48$ & $4,6,8,10,12$ & $62350$ \\
$1/6$ & $0.0463597$ & $0.823449$ & $36^2 \times 72$ & $6,12,18$ & $36440$ \\
$1/8$ & $0.0463597$ & $0.823449$ & $48^2 \times 96$ & $4,6,8,12,16,20,24$ & $29380$ \\
$1/12$ & $0.0463597$ & $0.823449$ & $72^2 \times 144$ & $6,12,18,24,30,36$ & $8780$ \\
$1/16$ & $0.0463597$ & $0.823449$ & $96^2 \times 192$ & $4,8,12,16,24,32,40$ & $3760$ \\
$1/24$ & $0.0463597$ & $0.823449$ & $144^2 \times 288$ & $6,12,18,24$ & $460$ \\
$1/32$ & $0.0463597$ & $0.823449$ & $192^2 \times 384$ & $4,8,16,24,32$ & $300$ \\
\hline
$1/4$ & $0.0178626$ & $1.64668$ & $24^2 \times 48$ & $4,6,8,10,12$ & $62340$ \\
$1/6$ & $0.0178626$ & $1.64668$ & $36^2 \times 72$ & $6,12,18$ & $36500$ \\
$1/8$ & $0.0178626$ & $1.64668$ & $48^2 \times 96$ & $4,6,8,12,16,20,24$ & $29580$ \\
$1/12$ & $0.0178626$ & $1.64668$ & $72^2 \times 144$ & $6,12,18,24,30,36$ & $8740$ \\
$1/16$ & $0.0178626$ & $1.64668$ & $96^2 \times 192$ & $4,8,12,16,24,32,40$ & $3700$ \\
$1/24$ & $0.0178626$ & $1.64668$ & $144^2 \times 288$ & $6,12,18,24$ & $480$ \\
$1/32$ & $0.0178626$ & $1.64668$ & $192^2 \times 384$ & $4,8,16,24,32$ & $300$ \\
\hline
\end{tabular} }
\caption{Parameters for all EQCD multi-level simulations.}
\label{sim_params}
\end{table}

\FloatBarrier
\bibliographystyle{unsrt}
\bibliography{references}

\end{document}